\documentclass[11pt]{article}

\usepackage{amsmath,amsfonts,amssymb}

\usepackage[numbers]{natbib}
\usepackage{graphicx}

\usepackage{hyperref}
\hypersetup{
   colorlinks   =  true,
    linkcolor    = cyan,
    citecolor    = red,
     urlcolor	=magenta,     
}

\newcommand\be{\begin{equation}}
\newcommand\ee{\end{equation}}
\newcommand\bea{\begin{eqnarray}}
\newcommand\eea{\end{eqnarray}}

   \newcommand\dd{{\rm d}}

\DeclareMathOperator\spn{span}

\begin{document}

\begin{center}
\baselineskip 24 pt {\LARGE \bf  
Noncommutative spaces of worldlines} 
\end{center}

\bigskip 
\medskip

\begin{center}

{\sc Angel Ballesteros, Ivan Gutierrez-Sagredo, Francisco J. Herranz}

{Departamento de F\'isica, Universidad de Burgos, 
09001 Burgos, Spain}

e-mail: {\href{mailto:angelb@ubu.es}{angelb@ubu.es}, \href{mailto:igsagredo@ubu.es}{igsagredo@ubu.es}, \href{mailto:fjherranz@ubu.es}{fjherranz@ubu.es}}

\end{center}

\begin{abstract}
The space of time-like geodesics on Minkowski spacetime is constructed as a coset space of the Poincar\'e group in (3+1) dimensions with respect to the stabilizer of a worldline. When this homogeneous space is endowed with a Poisson homogeneous structure compatible with a given Poisson-Lie Poincar\'e group, the quantization of this Poisson bracket gives rise to a noncommutative space of worldlines with quantum group invariance. As an oustanding example, the Poisson homogeneous space of worldlines coming from the $\kappa$-Poincar\'e deformation is explicitly constructed, and shown to define a symplectic structure on the space of worldlines. Therefore, the quantum space of $\kappa$-Poincar\'e worldlines is just the direct product of three Heisenberg-Weyl algebras in which the parameter $\kappa^{-1}$ plays the very same role as the Planck constant $\hbar$ in quantum mechanics. In this way, noncommutative spaces of worldlines are shown to provide a new suitable and fully explicit arena for the description of quantum observers with quantum group symmetry.
\end{abstract}


\bigskip

\noindent
KEYWORDS:  

\noindent time-like worldlines, quantum groups, Minkowski spacetime, Poisson homogeneous spaces, kappa-deformation, non-commutative spaces, quantum observers

%%%%%%%%%%%%%%%%%%%%%%%%%%%%%%%%%%%%%%%%%%%%%%%%%%%

\section{Introduction}

It is widely assumed that the spacetime underlying a fundamental theory of quantum gravity should include some noncommutative structure (see for instance~\cite{Snyder1947, DFR1994, MW1998, Szabo2003, FL2006}) that generates the Planck scale uncertainty relations describing the minimum length phenomena arising in the interplay between gravity and quantum theory~\cite{Garay1995}. In this context, a mathematically self-consistent approach to such `quantum geometry' consists in the construction of noncommutative (`quantum') spacetimes  which are invariant under quantum Poincar\'e or (A)dS groups (see~\cite{LRNT1991,LNR1992fieldtheory,MR1994,BHOS1995nullplane,Zakrzewski1997, BRH2003minkowskian,BLT2016unified, BLT2016unifiedaddendum, MS2018constraints, BM2018extended} and references therein), which are defined as both noncommutative and noncocommutative Hopf algebra deformations~\cite{Majid1988,ChariPressley1994,Majid1995Book} of the corresponding classical kinematical groups. In this way, several proposals for the analysis of the role played by quantum Lorentzian symmetries both on momentum space (see~\cite{MajidCMS,Kowalski-Glikman2013living,GM2013relativekappa,BGGH2017curvedplb,BGGH2018cms31} and references therein) and on phase space~\cite{LSW2015hopfalgebroids,LMMPW2018algebroid} have been presented in connection with deformed or doubly special relativity (DSR) theories \cite{Amelino-Camelia2001testable,Kowalski-Glikman2001, Amelino-Camelia2002planckian,  MS2002,LN2003versus,BRH2003newdoubly, FKS2004gravity, ASS2004,Amelino-Camelia2010symmetry} where generalizations of special relativity with a new invariant quantity additional to the speed of light and related with the Planck scale were introduced. Both in DSR theories and in the relative locality approach~\cite{AFKS2011deepening,AFKS2011principle,AAKRG2012relativelocality} the focus  is switched from spacetime to momentum space, under the observation that the only quantities that can truly be  measured are momenta and energy of particles.

The aim of this work is to propose noncommutative spaces of time-like worldlines as a new feasible alternative in order to describe the Planck scale effects encoded under quantum group symmetry, and to show that these noncommutative spaces of oriented time-like geodesics can be explicitly constructed in a systematic way. This can be achieved in a rigorous and completely general manner through the approach here presented, where we will introduce the `quantum' structure of the homogeneous space of worldlines corresponding to Lorentzian spacetimes as quantizations of Poisson homogeneous spaces of Lorentzian groups. Since time-like worldlines can be identified with inertial observers, this can be thought of as a first  --to the best of our knowledge-- explicit model of quantum observers with quantum group symmetry. 

Indeed, the construction here presented contains a main simplification, since we are assuming that the space of oriented time-like geodesics is also a homogeneous space, which will be the case for maximally symmetric Lorentzian spacetimes, namely the Minkowski case here studied and the (A)dS ones (see~\cite{HS1997phasespaces}). Recall that for a generic spacetime,
i.e.~a smooth manifold endowed with a pseudo-Riemannian metric, the space of oriented geodesics is a quite complicated object. In fact, it is a topological space but not necessarily Hausdorff, and even when this is the case the topological manifold could not admit a  smoothable atlas and henceforth could not be a smooth manifold. These problems have been previously considered in the literature (see, for instance,~\cite{Low1989} and~\cite{BP1991geodesics}, where the space of null geodesics is described). For manifolds whose geodesics are closed many results are known (see \cite{Besse1978bookgeodesics}).

Our construction will make heavy use of this homogeneous space assumption, since this will allow the construction of the noncommutative version of such spaces as quantum homogeneous spaces invariant under the associated quantum isometry groups. Moreover, it is well-known that most of the structures that can be defined on classical homogeneous spaces of geodesics can be inherited from their associated motion groups. For instance, in~\cite{AGK2011} all symplectic, complex and metric structures were described, and we recall that in~\cite{HS1997phasespaces} all homogeneous spaces of worldlines corresponding to kinematical groups were studied in detail, including the pseudo-Riemannian metrics defined on them, and in~\cite{BRH2017} the (2+1) Lorentzian spaces of worldlines were considered. In particular, for the Poincar\'e case it was found that an invariant foliation exists in the space of worldlines   and that the resulting homogeneous space is of negative curvature (see~\cite{HS1997phasespaces} for details). This result provides a neat geometrical description of the hyperbolic nature of the space of velocities in special relativity, and all these classical geometric notions should admit some rigorous generalization to the quantum (noncommutative) setting.

We stress that   the construction here presented is fully general and explicit, thus being amenable to be applied to any (coisotropic, as we will explain in the sequel) quantum deformation of the Poincar\'e and (A)dS kinematical groups. Nevertheless, for the sake of brevity we will restrict here to the case with vanishing cosmological constant, and
we will explicitly construct and analyse the noncommutative space of worldlines associated to the $\kappa$-Poincar\'e quantum group, whose relevance is outstanding (see~\cite{LRNT1991, LNR1992fieldtheory, MR1994, GM2013relativekappa, LN2003versus} and references therein).

In the next Section we review the construction of Minkowski spacetime and its space of oriented geodesics as coset spaces of the Poincar\'e group, by emphasizing the relevance of choosing appropriate (and different) sets of local coordinates on the group for the construction of each of these two spaces. In Section 3 we recall how the $\kappa$-Minkowski noncommutative spacetime arises as the quantization of the Poisson Minkowski spacetime associated with the $\kappa$-Poincar\'e Poisson-Lie structure, which is provided by the $\kappa$-Poincar\'e classical $r$-matrix. Section 4 mimics the same approach in order to construct the noncommutative $\kappa$-Poincar\'e space of worldlines as the quantization of the Poisson homogeneous spacetime of worldlines associated to the same $\kappa$-Poincar\'e Poisson-Lie group, which however has to be expressed in terms of a different set of local coordinates. As the most relevant result, we prove in  Section 5  that the non-trivial Poisson structure on the space of worldlines so obtained is just a canonical symplectic structure. This means that fuzziness can be expressed on the space of worldlines in a precise mathematical way, since
$\kappa$-Poincar\'e quantum observers can be described as the quantization of a canonical phase space of worldline coordinates in which the Planck constant $\hbar$ has been replaced by the quantum deformation parameter $\kappa^{-1}$. A final discussion Section including some open problems closes the paper.

%%%%%%%%%%%%%%%%%%%%%%%%%%%%%%%%%%%%%%%%%%%%%%%%%%%%%%%%%%

\section{The space of time-like geodesics as a coset space}

In this section we firstly recall the classical construction of the  (3+1)-dimensional  Minkowski spacetime as a coset space $\mathcal{M}=G/L$ of the Poincar\'e group $G$ with respect to the isotropy Lorentz subgroup $L$. In this way we obtain a maximally symmetric homogeneous space which can be   identified straightforwardly with Minkowski spacetime. Afterwards, we use the very same approach in order to construct the  6-dimensional  space of time-like oriented geodesics of Minkowski spacetime, just considering the coset $\mathcal{W}=G/H$, where now $H$ is the isotropy subgroup of a time-like geodesic that will be described in the sequel. This space inherits its main geometrical structures from the Poincar\'e group, as described in~\cite{HS1997phasespaces,AGK2011}. 

Let us consider the Poincar\'e Lie algebra $\mathfrak{g}=\mathfrak{p}(3+1) \equiv \mathfrak{so}(3,1) \ltimes \mathbb{R}^3$, which generates the (3+1) Poincar\'e group $G = P(3+1)$. In the kinematical basis $\{P_0,P_a, K_a, J_a\}$  $(a=1,2,3)$ of generators of time translation, space translations, boosts and rotations, respectively, the commutation rules for $\mathfrak{p}(3+1)$ read
\be
\begin{array}{lll}
[J_a,J_b]=\epsilon_{abc}J_c ,& \quad [J_a,P_b]=\epsilon_{abc}P_c , &\quad
[J_a,K_b]=\epsilon_{abc}K_c , \\[2pt]
\displaystyle{
  [K_a,P_0]=P_a  } , &\quad\displaystyle{[K_a,P_b]=\delta_{ab} P_0} ,    &\quad\displaystyle{[K_a,K_b]=-\epsilon_{abc} J_c} , 
\\[2pt][P_0,P_a]=0 , &\quad   [P_a,P_b]=0 , &\quad[P_0,J_a]=0  ,
\end{array}
\label{eq:ads_Liealg3+1}
\ee
where sum over repeated indices is assumed, latin indices run from 1 to 3, and the speed of light $c=1$ is assumed. In the rest of the paper the same conventions are followed, together with greek indexes running from 0 to 3  and denoting  3-vectors by $\mathbf{v}=(v^1,v^2,v^3)$ and 4-vectors by $\bar{v}=(v^0,v^1,v^2,v^3)$. Consider the Lie subalgebras of $\mathfrak{g}$ given by 
\be
\mathfrak{l} = \spn\{ K_a, J_a \}, \qquad \mathfrak{h} = \spn\{ P_0, J_a \},
\label{sub}
\ee 
corresponding to the Lie subgroups $L$ and $H$ of $G$, respectively. Geometrically, these Lie subgroups are  the stabilizers of the origin of Minkowski spacetime (an event)  and  that of the space of worldlines (a time-like oriented geodesic).

 The first non-trivial task is to find appropriate descriptions of both the spacetime $\mathcal{M}$ and the space of worldlines $\mathcal{W}$ in such a way that the projection of Poisson-Lie structures on the Poincar\'e group $G$ can be easily described in a canonical way. This problem turns out to be equivalent to the exponentiation of a faithful representation of the Lie algebra in two (different) and carefully chosen orders.

Explicitly, let us consider a faithful representation $\rho : \mathfrak{p}(3+1)\rightarrow \text{End}(\mathbb R ^5)$ such that a generic element $X$ of the Lie algebra $\mathfrak{p}(3+1)$ is given by:
\begin{equation}
\label{eq:repG}
\rho(X)=   x^\alpha \rho(P_\alpha)  +  \xi^a \rho(K_a) +  \theta^a \rho(J_a) =\left(\begin{array}{ccccc}
0&0&0&0&0\cr 
x^0 &0&\xi^1&\xi^2&\xi^3\cr 
x^1 &\xi^1&0&-\theta^3&\theta^2\cr 
x^2 &\xi^2&\theta^3&0&-\theta^1\cr 
x^3 &\xi^3&-\theta^2&\theta^1&0
\end{array}\right) .
\end{equation}
In order to construct the (3+1)-dimensional  Minkowski spacetime $\mathcal{M}$ as a coset space, we parametrize an element of the Poincar\'e group $G=P(3+1)$ in the form
\begin{align}
\begin{split}
\label{eq:Gm}
&G_\mathcal{M}= \exp{x^0 \rho(P_0)} \exp{x^1 \rho(P_1)} \exp{x^2 \rho(P_2)} \exp{x^3 \rho(P_3)} \\
&\qquad\quad\times \exp{\xi^1 \rho(K_1)} \exp{\xi^2 \rho(K_2)} \exp{\xi^3 \rho(K_3)}
 \exp{\theta^1 \rho(J_1)} \exp{\theta^2 \rho(J_2)} \exp{\theta^3 \rho(J_3)} \, ,
\end{split}
\end{align}
and the Lorentz subgroup $L$ is parametrized by
\begin{align}
\begin{split}
\label{eq:Lm}
&L= \exp{\xi^1 \rho(K_1)} \exp{\xi^2 \rho(K_2)} \exp{\xi^3 \rho(K_3)} \exp{\theta^1 \rho(J_1)} \exp{\theta^2 \rho(J_2)} \exp{\theta^3 \rho(J_3)} .
\end{split}
\end{align}
In this way the  coordinates $x^\alpha$ can be understood as Minkowski spacetime coordinates in the standard way, and the usual   spacetime metric onto $\mathcal{M}$
\begin{equation}
\dd s^2= (\dd x^0)^2- (\dd x^1)^2- (\dd x^2)^2- (\dd x^3)^2 ,
\label{Mmetric}
\end{equation}
arises.
Note that these coordinates are specially adapted to $\mathcal{M}$ because they come from a splitting of the Poincar\'e group given by $G_{\mathcal{M}}=T \cdot L$, where $T$ is the translations sector and $L$ is the Lorentz one, which locally is always valid. Hence   a Poincar\'e group element takes the standard form
\begin{align}
\begin{split}
\label{eq:Gm_split}
&G_\mathcal{M} =  \left(
\begin{array}{cc}
1 & \bar{0}  \\
\bar{x}^T & \mathbf{\Lambda} \\
\end{array}\right) ,
\end{split}
\end{align}
where   $\mathbf{\Lambda}$ is the $4 \times 4$ matrix representation of an element of the Lorentz subgroup.

While the previous construction is well-known, the corresponding one for the space of worldlines is more subtle, due to the fact that the coordinates $x^a$ and $\xi^a$ associated with space translations and boosts do not define a set of coordinates in the coset space $\mathcal{W}=G/H$, because they are not well-defined functions on this space. Recall that for a function to be well-defined on the coset space $\mathcal{W}=G/H$ it must be independent of the representative chosen. For instance, for $w$ being a well-defined set of six coordinates on $\mathcal{W}$ they must verify that $w(gh)=w(gh')$ for all $g \in G$ and $h,h' \in H$. A direct computation shows that this is not the case for $x^a$ and $\xi^a$ when taking the group law arising from the exponentiation~\eqref{eq:Gm} and considering the worldlines coset $\mathcal{W}=G/H$ (of course, $x^\alpha$ do define a proper set of coordinates for the spacetime coset $\mathcal{M}=G/L$).

Therefore, in order to construct the 6-dimensional  space of worldlines as the coset space $\mathcal{W}=G/H$ we need to parametrize $G$ by using a different ordering, which turns out to be
\begin{align}
\begin{split}
\label{eq:Gw}
&G_\mathcal{W}= \exp{\eta^1 \rho(K_1)} \exp{y^1 \rho(P_1)} \exp{\eta^2 \rho(K_2)} \exp{y^2 \rho(P_2)} \exp{\eta^3 \rho(K_3)} \exp{y^3 \rho(P_3)}  \\
& \qquad\qquad\times \exp{\phi^1 \rho(J_1)} \exp{\phi^2 \rho(J_2)} \exp{\phi^3 \rho(J_3)} \exp{y^0 \rho(P_0)},
\end{split}
\end{align}
where the stabilizer $H$ of a worldline passing through the origin of $\mathcal{W}$  is parametrized as 
\begin{align}
\begin{split}
\label{eq:Hw}
&H= \exp{\phi^1 \rho(J_1)} \exp{\phi^2 \rho(J_2)} \exp{\phi^3 \rho(J_3)} \exp{y^0 \rho(P_0)} .
\end{split}
\end{align}
In this way it is straightforward to check that we have well-defined coordinates $y^a$ and $\eta^a$ on $\mathcal{W}=G/H$, as they are invariant by right multiplication of an element of $H$. These coordinates correspond to the splitting of the Poincar\'e group as $G_\mathcal{W} = D \cdot H$, and the space of worldlines is just parametrized by the local coordinates dual to the generators of $D$. Now the group element has the form 
\begin{align}
\begin{split}
\label{eq:Gw_split}
&G_\mathcal{W} =  \left(
\begin{array}{cc}
1 & \bar{0}  \\
{\bar{f}}^T & \mathbf{\Lambda} \\
\end{array}\right),
\end{split}
\end{align}
where $\mathbf{\Lambda}$ is the same matrix as in~\eqref{eq:Gm_split} and $\bar{f}$ is formed by the   functions given by
\begin{align}
\begin{split}
\label{eq:falpha}
&f^0 (y^\alpha, \eta^a) = y^1 \sinh \eta^1 + \cosh \eta^1 \left( y^2 \sinh \eta^2 + \cosh \eta^2 (y^0 \cosh \eta^3 + y^3 \sinh \eta^3) \right) ,\\
&f^1 (y^\alpha, \eta^a) = y^1 \cosh \eta^1 + \sinh \eta^1 \left( y^2 \sinh \eta^2 + \cosh \eta^2 (y^0 \cosh \eta^3 + y^3 \sinh \eta^3) \right) ,\\
&f^2 (y^\alpha, \eta^a) = y^2 \cosh \eta^2 + \sinh \eta^2 (y^0 \cosh \eta^3 + y^3 \sinh \eta^3), \\
&f^3 (y^\alpha, \eta^a) = y^0 \sinh \eta^3 + y^3 \cosh \eta^3 \, .
\end{split}
\end{align}
We stress that the previous construction allows us to obtain the explicit relationships among the local coordinates of the Poincar\'e group $G$ in both parametrizations, namely 
\be
\bar{x} = \bar{f} (y^\alpha, \eta^a) , \qquad \xi^a = \eta^a ,\qquad  \theta^a  =  \phi^a .
\label{xf}
\ee
Note that the position coordinates $x^a\equiv f^a$ on the Minkowski spacetime cannot be naively  identified with the `position' coordinates  $y^a$ on the space of worldlines,  since they only coincide when all rapidities vanish (i.e.~for an observer at rest). Moreover, in the representation~\eqref{eq:Gw_split} the action of the Poincar\'e group on the space of worldlines is not linear.

 The space  of worldlines $\mathcal{W}$  has also a metric structure which   is rather different from the flat Lorentzian metric on the  Minkowskian spacetime $\mathcal{M}$ (\ref{Mmetric}). In particular, the metric on  $\mathcal{W}$  is degenerate, and an invariant foliation under the Poincar\'e group action arises in such a manner  that  a `subsidiary' metric   restricted to each leaf of the foliation has to be considered~\cite{HS1997phasespaces}. It can be shown that in terms of the coordinates $y^a$ and $\eta^a$ the  degenerate `main' metric $g^{(1)}$ on $\mathcal{W}$ has a line element 
 \be
 {\rm d} s_{(1)}^2=(\cosh\eta ^2)^2 (\cosh\eta^3)^2(\dd \eta^1)^2+ (\cosh\eta^3)^2 (\dd \eta^2)^2+( \dd \eta^3 )^2.
 \label{metric1}
 \ee
This is a   Riemannian metric of negative constant curvature (whose value is just  $-1/c^2$) which only involves   rapidities, and  thus     provides the relative rapidity between two free motions.  
This,   in turn, shows that the three-velocity space is hyperbolic. The invariant foliation is determined by a uniform motion with $\boldsymbol{\eta}= {\boldsymbol{\eta}}_0=$\,constant and the `subsidiary' metric $g^{(2)}$ defined on each leaf  reads
 \be
 {\rm d} s_{(2)}^2=( \dd y^1)^2+  ( \dd y^2)^2+( \dd y^3)^2 ,\qquad   \boldsymbol{\eta}= {\boldsymbol{\eta}}_0 ,
 \label{metric2}
 \ee
that is, each leaf is isometric to the three-dimensional Euclidean space.  We also point out that  in the  three-velocity space the geodesic distance $\chi$ corresponding to the relative speed from an observer at rest and one with a uniform motion with rapidity $ \boldsymbol{\eta}$ is given by
 \be
\cosh\chi=\cosh\eta ^1 \cosh\eta ^2 \cosh\eta^3.
 \label{distance}
 \ee
Finally, notice that in the low rapidity regime (i.e.~take $c\to \infty$), the expressions (\ref{metric1}) and (\ref{distance}) reduce to the usual ones for velocities in classical mechanics
\be
{\rm d} s_{(1)}^2= (\dd \eta^1)^2+  (\dd \eta^2)^2+( \dd \eta^3 )^2,\qquad \chi^2=  (  \eta^1)^2+  (  \eta^2)^2+(   \eta^3 )^2 .
\label{classical}
\ee

%%%%%%%%%%%%%%%%%%%%%%%%%%%%%%%%%%%%%%%%%%%%

\section{$\kappa$-Minkowski noncommutative spacetime}

For the sake of clarity and self-consistency we now recall the well-known construction of the $\kappa$-Minkowski noncommutative spacetime as the quantization of a coisotropic Poisson homogenous structure on $\mathcal{M}$ obtained from the $\kappa$-Poincar\'e Poisson-Lie structure on the Poincar\'e group (which has to be parametrized in the form $G_\mathcal{M}$ (\ref{eq:Gm_split})) through canonical projection.

The $\kappa$-Poincar\'e Lie bialgebra is a coboundary Lie bialgebra generated by the following $r$-matrix
\be
\label{eq:r_kappapoincare}
r=\frac{1}{\kappa} (K_1 \wedge P_1 + K_2 \wedge P_2 + K_3 \wedge P_3) .
\ee
This means that the associated cocommutator can be directly computed as $\delta(X) = [X \otimes 1 + 1 \otimes X,r], \; \forall X \in \mathfrak{g}$, and reads 
 \begin{align}
\begin{split}
\label{eq:cocom_kappapoincare}
& \delta(P_0) = \delta(J_a) = 0 ,\\
& \delta(P_a) = \frac{1}{\kappa} P_a \wedge P_0 ,\\
& \delta(K_1) = \frac{1}{\kappa} (K_1 \wedge P_0 + J_2 \wedge P_3 - J_3 \wedge P_2), \\
& \delta(K_2) = \frac{1}{\kappa}  (K_2 \wedge P_0 + J_3 \wedge P_1 - J_1 \wedge P_3), \\
& \delta(K_3) = \frac{1}{\kappa}  (K_3 \wedge P_0 + J_1 \wedge P_2 - J_2 \wedge P_1) .
\end{split}
\end{align}
Coboundary Lie bialgebras are the tangent counterpart of coboundary Poisson-Lie groups~\cite{ChariPressley1994}, where the Poisson structure on $G$ is given by the so-called Sklyanin bracket
\begin{align}
\begin{split}
\label{eq:sklyanin}
&\{f,g\}=r^{ij}\left( X^L_i f \, X^L_j g - X^R_i f \, X^R_j g \right),\qquad f,g \in \mathcal C (G),
\end{split}
\end{align} 
such that    $X^L_i$ and $ X^R_j$ are   left- and right-invariant vector fields  defined by
\begin{align}
\label{eq:ivf}
X^L_i f(h)&=\frac{\dd}{\dd t}\biggr\rvert _{t=0} f\left(h\, {\rm e}^{t T_i}\right),  \qquad  X^R_i f(h)=\frac{\dd}{\dd t}\biggr\rvert _{t=0} f\left({\rm e}^{t T_i} h\right),
\end{align}
 where $f \in \mathcal C (G)$, $h \in G$ and $T_i \in \mathfrak g$. 

The Poisson brackets defining the $\kappa$-Minkowski spacetime are just the ones providing the Poisson homogeneous structure on $\mathcal{M}$ associated to the Poisson-Lie structure on the Poincar\'e group defined by the $r$-matrix~\eqref{eq:r_kappapoincare}. In order to obtain them explicitly, it suffices to realize that the $\kappa$-Poincar\'e Lie bialgebra~\eqref{eq:cocom_kappapoincare} is a coisotropic Lie bialgebra  with respect to the Lorentz subgroup, since it fulfills the condition (see~\cite{BMN2017homogeneous} and references therein)
\be
\delta (\mathfrak{l}) \subset \mathfrak{l} \wedge \mathfrak{g} \, ,
\label{coisotropicc}
\ee
where $\mathfrak{l}$ (\ref{sub})  is the Lie algebra of the Lorentz subgroup $L$.   In that case, it turns out that the homogeneous Poisson structure for the coset space $\mathcal{M}=G/L$ is just the canonical projection of the Poisson-Lie bracket~\eqref{eq:sklyanin} to the Minkowski space coordinates $x^\alpha$, which is tantamount to say  that  the (Poisson) $\kappa$-Minkowski spacetime is given by the Sklyanin bracket (\ref{eq:sklyanin}) between the Minkowski coordinates, which can be straightforwardly computed and reads
\begin{align}
\begin{split}
\label{eq:PoissonkappaMinkowski}
&\{x^0,x^a\}=- \frac{1}{\kappa} \, x^a, \qquad  \{x^a,x^b\}= 0\, .
\end{split}
\end{align} 

This bracket can be quantized just by replacing the Poisson brackets by commutators, since no ordering problems arise due to the linear nature of the Poisson structure \cite{Maslanka1993}\begin{align}
\begin{split}
\label{eq:conmkappaMinkowski}
&[\hat x^0,\hat x^a]=- \frac{1}{\kappa} \, \hat x^a, \qquad \qquad [\hat x^a,\hat x^b]= 0\, ,
\end{split}
\end{align} 
hence $\hat x^\alpha$ will be the noncommutative coordinates on this quantum ($\kappa$-Minkowski) spacetime. These relations are directly linked through Hopf algebra duality to the quantum $\kappa$-Poincar\'e Hopf algebra, which in the bicrossproduct basis~\cite{MR1994} is given by the deformed commutation rules
\be
[K_a,P_b]=\delta_{ab} \left( \frac{\kappa}{2} \left(1-{\rm e}^{-2   P_0/\kappa} \right) + \frac{1}{2 \kappa}\, \textbf{P}^2 \right) -  \frac{1}{\kappa}\, P_a P_b \, ,
\ee
where $\textbf{P}^2 = P_1^2 + P_2^2 +P_3^2$, and
the rest of commutators being the undeformed ones \eqref{eq:ads_Liealg3+1}. The corresponding deformed coproduct is given by
\begin{align}
\begin{split}
\label{eq:coproductkappaPoincare}
&\Delta (P_0) = P_0 \otimes 1 + 1 \otimes P_0, \\
&\Delta (J_a) = J_a \otimes 1 + 1 \otimes J_a, \\
&\Delta (P_a) = P_a \otimes 1 +{\rm  e}^{-  P_0 /\kappa} \otimes P_a, \\
&\Delta (K_a) = K_a \otimes 1 + {\rm e}^{-  P_0 /\kappa} \otimes K_a +  \frac{1}{\kappa}\, \epsilon_{abc} P_b \otimes J_c\, .
\end{split}
\end{align}
Recall that the skew-symmetric part of the first-order in $\kappa^{-1}$ of the coproduct~\eqref{eq:coproductkappaPoincare} provides the cocommutator~\eqref{eq:cocom_kappapoincare}, where quantum spacetime coordinates are dual generators to the deformed translations sector through the pairing $
\langle \hat x^\alpha, P_\beta \rangle= \delta_\beta^\alpha$.

%%%%%%%%%%%%%%%%%%%%%%%%%%%%%%%%%%%%%%%%%%%%%%%%%%%%%

\section{$\kappa$-Poincar\'e homogeneous space of worldlines}

The main result presented in this paper consists is showing that the noncommutativity in the space of worldlines induced by the $\kappa$-deformation of Poincar\'e symmetries can be obtained by mimicking the previous construction of the Poisson homogeneous Minkowski spacetime, but taking into account the appropriate isotropy subgroup of worldlines. 

As a first step, we have to check whether the $\kappa$-Poincar\'e Lie bialgebra structure~\eqref{eq:cocom_kappapoincare} is coisotropic with respect to the Lie subalgebra of the isotropy subgroup of time-like worldlines $\mathfrak{h} = \spn\{ P_0, J_a \}$ (\ref{sub}). This is indeed the case, since we have that $\delta(P_0) = \delta(J_a) = 0$ and the coisotropy condition~\eqref{coisotropicc} is trivially satisfied. Furthermore, in this case the stronger Poisson-subgroup condition~\cite{BMN2017homogeneous} $\delta(\mathfrak{h}) \subset \mathfrak{h} \wedge \mathfrak{h}$ is also trivially fulfilled. We stress that this does not occur for the $\kappa$-Minkowski isotropy subalgebra  $\mathfrak{l}$, and implies that after quantization the isotropy subgroup $\mathfrak{h}$ is promoted to a Hopf subalgebra (again, this is not the case for the Lorentz sector of the $\kappa$-deformation). 

This fact provides a first signature that  the $\kappa$-deformation is more naturally realized on the space of worldlines than on Minkowski spacetime. As a consequence, the coisotropy condition guarantees that the homogeneous Poisson structure on the space of worldlines $\mathcal{W}$ can be obtained as a canonical projection from the coboundary Poisson structure on $G$ induced by the $r$-matrix~\eqref{eq:r_kappapoincare}. This is directly connected with the precise ordering~\eqref{eq:Gw} chosen for the construction of $G_\mathcal{W}$ (\ref{eq:Gw_split}), which is the one that provides the appropriate description of the projected Poisson structure onto $\mathcal{W}$. 

Indeed, left- and right-invariant vector fields~\eqref{eq:ivf} for $G_\mathcal{W}$ have to be computed (we omit their explicit expressions for the sake of brevity), and the Sklyanin bracket~\eqref{eq:sklyanin} for the $\kappa$-Poincar\'e $r$-matrix~\eqref{eq:r_kappapoincare} has to be written in terms of such vector fields expressed in terms of the local Poincar\'e coordinates $\{y^a,\eta^a,\phi^a,y^0\}$. With all these ingredients at hand, the explicit form of the $\kappa$-Poincar\'e Poisson homogeneous space of  worldlines is just given by (the canonical projection of) the Skyanin bracket for the coordinates  $y^a$ and $\eta^a$  of the space $\mathcal{W}$, and reads
\begin{align}
\begin{split}
\label{eq:pois_y_eta}
&\{y^1,y^2\}= \frac 1{\kappa} \left(y^2 \sinh \eta^1 - \frac{y^1 \tanh \eta^2}{\cosh \eta^3} \right) ,\\
&\{y^1,y^3\}=  \frac 1{\kappa}    \left(y^3 \sinh \eta^1 - y^1 \tanh \eta^3 \right) , \\
&\{y^2,y^3\}=   \frac 1{\kappa}    \left(y^3 \cosh \eta^1 \sinh \eta^2 - y^2 \tanh \eta^3\right) ,\\
&\{y^1,\eta^1\}=  \frac 1{\kappa}  \frac{ \left(  \cosh \eta^1  \cosh \eta^2 \cosh \eta^3-1 \right)}{\cosh \eta^2 \cosh \eta^3},\\
&\{y^2,\eta^2\}=  \frac 1{\kappa}   \frac{ \left( \cosh \eta^1 \cosh \eta^2\cosh \eta^3 -1\right)}{\cosh \eta^3}  ,\\
&\{y^3,\eta^3\}=   \frac 1{\kappa}  \left( \cosh \eta^1 \cosh \eta^2 \cosh \eta^3-1 \right) ,\\
& \{y^a,\eta^b\}= 0,\quad a\ne b,   \qquad  \{\eta^a,\eta^b\}=  0 \, .
\end{split}
\end{align}
These expressions for the Poisson version of the noncommutative space of worldlines show that this noncommutative space contains a commutative subalgebra of rapidities $\eta^a$, while the `position' worldline coordinates $y^a$ are noncommutative. The Poisson bracket between a given rapidity $\eta^a$ and its corresponding `position' $y^a$ does not vanish and depends on the geodesic distance function (\ref{distance}).

The structure of these Poisson brackets for $\mathcal{W}$ becomes more symmetric and manifestly spatially isotropic if they are expanded as power series in the coordinates of  $\mathcal{W}$ up to second-order, namely
\begin{align}
\begin{split}
\label{eq:pois_y_eta_ser}
&\{y^a,y^b\}=  \frac 1{\kappa}  \bigl(\eta^a y^b - \eta^b y^a \bigr) + \mathcal{O}(\bold{y},\boldsymbol{\eta})^3, \qquad    \{\eta^a,\eta^b\}= 0 ,\\
&\{y^a,\eta^b\}=  \delta_{ab} \, \frac{1}{2\kappa}\left( (\eta^1)^2 + (\eta^2)^2 + (\eta^3)^2 \right) + \mathcal{O}(\bold{y},\boldsymbol{\eta})^3 .\\
\end{split}
\end{align}
We stress that the quadratic terms in~\eqref{eq:pois_y_eta_ser} are the ones coming from the non-relativistic limit $c\to \infty$, since they are just the   angular momenta $(\eta^a y^b - \eta^b y^a)$ and the relative speed $\chi^2$ (\ref{classical}).   Note also  that the linearization of the brackets~\eqref{eq:pois_y_eta_ser} in terms of the local coordinates $y^a$ and $\eta^b$ vanish, in contradistinction to the $\kappa$-Minkowski spacetime~\eqref{eq:PoissonkappaMinkowski}, which is a purely linear bracket having no higher-order terms in the coordinates.

If we now compute the $\kappa$-Poincar\'e Sklyanin bracket relations between $y^0$ (which does not belong to $\mathcal{W}$) and the coordinates of the space of worldlines we obtain
\begin{equation}
\begin{split}
\label{eq:pois_y_eta_y0}
&\{y^0, y^1 \} = - \frac 1{\kappa} \left( y^1 - y^2\, \frac{\sinh \eta^1 \tanh \eta^2}{\cosh \eta^3} - y^3 \sinh \eta^1 \tanh \eta^3 \right) ,\\
&\{y^0, y^2 \} = - \frac 1{\kappa} \left( y^2 + y^1\, \frac{\sinh \eta^1 \tanh \eta^2}{\cosh \eta^3} - y^3 \cosh \eta^1 \sinh \eta^2 \tanh \eta^3  \right) ,\\
&\{y^0, y^3 \} = - \frac 1{\kappa} \left( y^3 + y^1 \sinh \eta^1 \tanh \eta^3 + y^2 \cosh \eta^1 \sinh \eta^2 \tanh \eta^3 \right) ,\\
&\{y^0, \eta^1 \} = - \frac 1{\kappa}\, \frac{\sinh \eta^1}{\cosh \eta^2 \cosh \eta^3} ,\\
&\{y^0, \eta^2 \} = - \frac 1{\kappa}\, \frac{\cosh \eta^1 \sinh \eta^2}{\cosh \eta^3} ,\\
&\{y^0, \eta^3 \} = - \frac 1{\kappa} \cosh \eta^1 \cosh \eta^2 \sinh \eta^3  ,
\end{split}
\end{equation}
which means that the smooth functions on $\mathcal{W}$ enlarged with $y^0$ ({\em i.e.} the coset $G/R$ where $R$ is the rotations subgroup) still define a Poisson subalgebra where $\mathcal{C}^\infty (\mathcal{W})$ is a non-Abelian ideal. All these properties are fully consistent with the transformation~\eqref{eq:falpha} which provides the Minkowski spacetime coordinates $x^\alpha$ in terms of the ones for $\mathcal{W}$ and $y^0$. Indeed, if we use the transformation~\eqref{eq:falpha} to compute the Poisson structure given by \eqref{eq:pois_y_eta} and \eqref{eq:pois_y_eta_y0} for the four Minkowski coordinates $x^\alpha\equiv f^\alpha$, we just obtain the defining relations of the Poisson version of the $\kappa$-Minkowski spacetime~\eqref{eq:PoissonkappaMinkowski}. This means that both the noncommutative $\kappa$-Minkowski spacetime and the noncommutative space of $\kappa$-Poincar\'e worldlines are just two realisations in two different geometric contexts of the very same noncommutative structure provided by the $\kappa$-deformation.

%%%%%%%%%%%%%%%%%%%%%%%%%%%%%%%%%%%%%%%%%%%%

\section{Quantum $\kappa$-Poincar\'e worldlines}

At this point, the space of quantum worldlines would be defined by the quantization of the Poisson algebra of worldline coordinates~\eqref{eq:pois_y_eta}, which can be obtained by substituting the Poisson brackets into commutators with exactly the same expressions~\eqref{eq:pois_y_eta}, but now in terms of the noncommutative quantum worldline coordinates $\hat y^a$ and $\hat \eta^a$. 
Indeed, no ordering ambiguities appear when the Poisson bracket~\eqref{eq:pois_y_eta} is transformed into a commutator: the quantum rapidities $\hat \eta^a$ commute, and this implies that the crossed commutators $[\hat y^a,\hat \eta^b]$ have no ordering problems. Finally, at the r.h.s.~of the commutators $[\hat y^a,\hat y^b]$, the $\hat y^a$ coordinates are always multiplied by a function depending on  the $\hat \eta^b$ coordinates  with $b \neq a$, which means that the latter commute with the former
and no quantization ambiguities do exist.

Moreover, the fact that the  Poisson brackets $\{y^a,\eta^b\}$ only involve the $\eta^a$ coordinates provides a natural ansatz for a new set of classical variables whose quantization is straightforward. Let us consider the following diffeomorphism
\begin{align}
\begin{split}
\label{eq:canonical_diffeomorphism}
& q^1 = \frac{\cosh \eta^2 \cosh \eta^3 }{\cosh \eta^1 \cosh \eta^2 \cosh \eta^3 - 1}\,y^1, \qquad   p^1=\eta^1,\\
& q^2 = \frac{\cosh \eta^3}{\cosh \eta^1 \cosh \eta^2 \cosh \eta^3 - 1}\, y^2, \qquad   p^2=\eta^2,\\
& q^3 = \frac{1}{\cosh \eta^1 \cosh \eta^2 \cosh \eta^3 - 1}\,y^3, \qquad   p^3=\eta^3,\\
\end{split}
\end{align}
which is well-defined whenever $(\eta^1, \eta^2, \eta^3) \neq (0,0,0)$ since its Jacobian determinant 
\be
\label{eq:jac_det}
|\bold{\text{J}} (\bold{y},\boldsymbol{\eta})|=\frac{\cosh \eta^2 (\cosh \eta^3)^2}{(\cosh \eta^1 \cosh \eta^2 \cosh \eta^3 - 1)^3} \, ,
\ee
is different from zero everywhere on such a domain (note again the presence of the  geodesic distance $\chi$ (\ref{distance}) in  all these expressions). In fact, in terms of the elements of $g^{(1)}$ \eqref{metric1} the expressions above take the simple form 
\begin{equation}
q^a =\frac{\sqrt{g_{aa}^{(1)}}}{\cosh \chi -1} y^a, \quad \quad |\bold{\text{J}} (\bold{y},\boldsymbol{\eta})|= \frac{\sqrt{\det g^{(1)}}}{(\cosh \chi -1)^3} \, ,
\end{equation}
where in this case sum over repeated indices should not be assumed. These expressions neatly shows the interconnection among the Poisson homogeneous structure underlying the quantum deformation and the (hyperbolic) geometry of the space of velocities of special relativity.

Surprisingly enough, in terms of these new coordinates on $\mathcal{W}$ the noncommutative (Poisson) algebra of the worldline coordinates turns out to be 
\begin{align}
\label{eq:pois_canonical}
\{q^a,q^b\}= \{p^a,p^b\}= 0, \qquad   \{q^a,p^b\}= \frac 1 {\kappa} \, \delta_{ab}  .
\end{align}
Obviously, the chosen notation $(q^a,p^a)$ for these new coordinates is not arbitrary, since what we have found is that the homogeneous Poisson structure induced by the $\kappa$-Poincar\'e $r$-matrix on the space or worldlines is just a symplectic structure on $\mathcal{W}$ (without the origin). Note that the diffeomorphism~\eqref{eq:canonical_diffeomorphism} is defined everywhere but in a point is a direct consequence of the fact that a Poisson-Lie group is never symplectic~\cite{ChariPressley1994}, because it always vanishes at the identity and this property descends to the quotient through canonical projection. Hence, outside the origin of $\mathcal W$ (the projection of the identity element to the coset space) we have obtained a symplectic form onto $\mathcal{W}$ given by 
\be
\label{eq:symplecticform}
\omega = \kappa \sum_{a=1}^3 \dd q^a \wedge \dd p^a \, ,
\ee
and in this way our new worldline coordinates  $(\bold{q},\bold{p})$  have a direct interpretation as canonical phase space coordinates. Notice that while the change $\frac{y^a}{\{y^a,\eta^a\}} \rightarrow q^a$ is clearly suggested by the precise form of the fundamental brackets \eqref{eq:pois_y_eta}, the fact that the new coordinates $q^a$ Poisson commute, which is essential for them in order to be Darboux coordinates, is a quite surprising result. Also, it is worth stressing that the symplectic structure~\eqref{eq:symplecticform} is the result of the $\kappa$-deformation of the Poincar\'e symmetry, and this explains why the deformation parameter $\kappa$ explicitly appears within the symplectic form. Note that taking the limit $\kappa \rightarrow \infty$ implies that the Poisson-Lie structure on the Poincar\'e group becomes the trivial one and the space of worldlines (\ref{eq:pois_y_eta}) and (\ref{eq:pois_canonical})  (as well as the spacetime~\eqref{eq:PoissonkappaMinkowski}) becomes commutative.

Obviously, the quantization of the symplectic algebra~\eqref{eq:pois_canonical} is straightforward in terms of canonical worldline position operators $\hat {\bold{q}}$ and their conjugate momenta $\hat {\bold{p}}$, and taking into account that     $\kappa^{-1}$  plays exactly the same role as the Planck constant. Therefore, all quantum gravity effects amenable to be described through the $\kappa$-deformation should be fully understood as a standard deformation-quantization on the space of worldlines with deformation parameter $\kappa^{-1}$, just in the same way as ordinary quantum mechanics arises as a deformation-quantization (with parameter $\hbar$) on the classical mechanical phase space.

%%%%%%%%%%%%%%%%%%%%%%%%%%%%%%%%%%%
\section{Discussion and open problems}

As a conclusion, we propose that quantum deformations of spaces of worldlines arising as quantisations of Poisson homogeneous spaces of kinematical groups should also be  considered as noncommutative spaces amenable to describe quantum gravity effects. We have shown that this is a completely general construction that can be applied to any quantum deformation provided that the coisotropy condition~\eqref{coisotropicc} of its associated Lie bialgebra with respect to the isotropy subalgebra of worldlines is fulfilled. In this way, a non-trivial Poisson homogenous  structure on the space of worldlines can be introduced, and from the latter the noncommutativity between worldline coordinates arises in a natural way. 
 
Furthermore, when this construction is applied to the $\kappa$-deformation of Poincar\'e symmetries, the quantum space of worldlines so obtained turns out to be isomorphic to three copies of the Heisenberg-Weyl algebra, where the constant $\kappa^{-1}$ plays the role of $\hbar$. This is a straightforward consequence of the symplectic structure of the space of worldlines induced by the $\kappa$-Poincar\'e Poisson-Lie structure. This result suggests that the $\kappa$-deformation is a natural one for the space of worldlines, an idea enforced by the fact that the isotropy subgroup of worldlines behaves as a Poisson-Lie subgroup under this deformation. Moreover, the fact that the $\kappa$-Poincar\'e $r$-matrix~\eqref{eq:r_kappapoincare} is connected in a natural way to a symplectic structure on the space of worldlines seems quite natural if we realize that the relativistic Newton-Wigner position operators~\cite{NW1971} are defined in terms of the Poincar\'e Lie algebra generators as
\be
Q_a=\frac{1}{2 P_0} \, K_a + K_a\, \frac{1}{2 P_0},
\label{eq:newtonwigner}
\ee
since in this way the canonical brackets $
[Q_a,P_b]=\delta_{ab}
$ 
are obtained in terms of the generators of the Poincar\'e Lie algebra~\eqref{eq:ads_Liealg3+1}. Therefore, the bivector
\be
B= Q_1 \wedge P_1 + Q_2 \wedge P_2 + Q_3 \wedge P_3 ,
\ee
should define a symplectic form under the appropriate realization, and if we substitute~\eqref{eq:newtonwigner} into this expression we get
\be
B= \left(\frac{1}{2 P_0} \, K_a + K_a\, \frac{1}{2 P_0} \right) \wedge P_a ,
\ee
which  implies that both the symplectic bivector $B$ and the $\kappa$-Poincar\'e $r$-matrix~\eqref{eq:r_kappapoincare} are closely related (recall that $P_0$ is one of the generators of the the stabilizer of the origin of the space of worldlines).

Consequently, the noncommutative spaces of worldlines seem to provide a privileged arena in order to explore the physical role of the $\kappa$-deformation. In particular, once the canonical coordinates have been found, noncommutativity in the space of worldlines could be rephrased in more physical terms as the impossibility of determining simultaneously and with infinite precision the six $(\bold{q},\bold{p})$ coordinates of a given worldline. In this respect, note that~\eqref{eq:canonical_diffeomorphism} implies that, before introducing the quantum deformation, the $\bold{p}$ coordinates are just the usual rapidities $\boldsymbol{\eta}$ and the `positions' $\bold{q}$ for a worldline are defined as the product of Poincar\'e coordinates $\bold{y}$ associated to translations with certain functions depending on $\boldsymbol{\eta}$. Indeed, the precise physical meaning of the coordinates~\eqref{eq:canonical_diffeomorphism} has to be studied in detail.

The construction here presented is fully general and can thus be applied to any other quantum deformation (provided it is coisotropic with respect to the isotropy subgroup of worldlines) of any kinematical group. This opens the path to several future investigations, and the first of them consists in the construction of the noncommutative space of worldlines associated with the $\kappa$-deformation of the (A)dS groups. This can be explicitly performed by generalizing the approach here presented to the case of nonvanishing cosmological constant $\Lambda$ through the approach to $\kappa$-(A)dS groups and spaces used in~\cite{BHOS1994global,BHM2014plb,BHMN2017kappa3+1}, in which $\Lambda$ is included as an explicit parameter whose $\Lambda\to 0$ limit provides the flat Poincar\'e case presented in this paper.

Finally, the model here introduced provides a purely kinematical framework for a sche\-matic theory of quantum (`noncommutative') free observers, and strongly suggests that if the quantum deformation is assumed to encode quantum gravity effects, the latter are simply reflected as a canonical Heisenberg-Weyl noncommutativity on the (phase) space of worldlines, which is the simplest possible algebraic framework to deal with. Work on all these lines is in progress and will be presented elsewhere.

%%%%%%%%%%%%%%%%%%%%%%%%%%%%%%%%%%%%%%%%%%%%%%%

\section*{Acknowledgements}

This work has been partially supported by Ministerio de Ciencia, Innovaci\'on y Universidades (Spain) under grant MTM2016-79639-P (AEI/FEDER, UE), by Junta de Castilla y Le\'on (Spain) under grant BU229P18 and by the Action MP1405 QSPACE from the European Cooperation in Science and Technology (COST). I.G-S. acknowledges a predoctoral grant from Junta de Castilla y Le\'on and the European Social Fund.

%%%%%%%%%%%%%%%%%%%%%%%%%%%%%%%%%%%%%%%%%%%%%%%

\small

%%%%%%%%%%%%%%BIBLIOGRAPHY%%%%%%%%%%%%%%%


\begin{thebibliography}{10}

\bibitem{Snyder1947}
H.~Snyder.
\newblock {Quantized Space-Time}.
\newblock {\em Phys. Rev.}, 71(1):38--41, 1947.
\newblock \href {http://dx.doi.org/10.1103/PhysRev.71.38}
  {\path{doi:10.1103/PhysRev.71.38}}.

\bibitem{DFR1994}
S.~Doplicher, K.~Fredenhagen, and J.~E. Roberts.
\newblock {Spacetime quantization induced by classical gravity}.
\newblock {\em Phys. Lett. B}, 331(1-2):39--44, jun 1994.
\newblock \href {http://dx.doi.org/10.1016/0370-2693(94)90940-7}
  {\path{doi:10.1016/0370-2693(94)90940-7}}.

\bibitem{MW1998}
H.-J. Matschull and M.~Welling.
\newblock {Quantum Mechanics of a Point Particle in (2+1)-dimensional gravity}.
\newblock {\em Class. Quantum Gravity}, 15:2981--3030, 1998.
\newblock \href {http://arxiv.org/abs/gr-qc/9708054v2}
  {\path{arXiv:gr-qc/9708054v2}}, \href
  {http://dx.doi.org/10.1088/0264-9381/15/10/008}
  {\path{doi:10.1088/0264-9381/15/10/008}}.

\bibitem{Szabo2003}
R.~J. Szabo.
\newblock {Quantum field theory on noncommutative spaces}.
\newblock {\em Phys. Rep.}, 378(4):207--299, 2003.
\newblock \href {http://arxiv.org/abs/hep-th/0109162}
  {\path{arXiv:hep-th/0109162}}, \href
  {http://dx.doi.org/10.1016/S0370-1573(03)00059-0}
  {\path{doi:10.1016/S0370-1573(03)00059-0}}.

\bibitem{FL2006}
L.~Freidel and E.~R. Livine.
\newblock {3D quantum gravity and effective noncommutative quantum field
  theory}.
\newblock {\em Phys. Rev. Lett.}, 96(22):221301, 2006.
\newblock \href {http://arxiv.org/abs/hep-th/0512113}
  {\path{arXiv:hep-th/0512113}}, \href
  {http://dx.doi.org/10.1103/PhysRevLett.96.221301}
  {\path{doi:10.1103/PhysRevLett.96.221301}}.

\bibitem{Garay1995}
L.~J. Garay.
\newblock {Quantum gravity and minimum length}.
\newblock {\em Int. J. Mod. Phys. A}, 10(2):145--165, 1995.
\newblock \href {http://arxiv.org/abs/gr-qc/9403008}
  {\path{arXiv:gr-qc/9403008}}, \href
  {http://dx.doi.org/10.1142/S0217751X95000085}
  {\path{doi:10.1142/S0217751X95000085}}.

\bibitem{LRNT1991}
J.~Lukierski, H.~Ruegg, A.~Nowicki, and V.~N. Tolstoy.
\newblock {q-deformation of Poincar{\'{e}} algebra}.
\newblock {\em Phys. Lett. B}, 264(3-4):331--338, 1991.
\newblock \href {http://dx.doi.org/10.1016/0370-2693(91)90358-W}
  {\path{doi:10.1016/0370-2693(91)90358-W}}.

\bibitem{LNR1992fieldtheory}
J.~Lukierski, A.~Nowicki, and H.~Ruegg.
\newblock {New quantum Poincar{\'{e}} algebra and $\kappa$-deformed field
  theory}.
\newblock {\em Phys. Lett. B}, 293(3-4):344--352, 1992.
\newblock \href {http://dx.doi.org/10.1016/0370-2693(92)90894-A}
  {\path{doi:10.1016/0370-2693(92)90894-A}}.

\bibitem{MR1994}
S.~Majid and H.~Ruegg.
\newblock {Bicrossproduct structure of $\kappa$-Poincar{\'{e}} group and
  non-commutative geometry}.
\newblock {\em Phys. Lett. B}, 334(3-4):348--354, 1994.
\newblock \href {http://dx.doi.org/10.1016/0370-2693(94)90699-8}
  {\path{doi:10.1016/0370-2693(94)90699-8}}.

\bibitem{BHOS1995nullplane}
A.~Ballesteros, F.~J. Herranz, M.~A. del Olmo, and M.~Santander.
\newblock {A new "null-plane" quantum Poincar{\'{e}} algebra}.
\newblock {\em Phys. Lett. B}, 351(1-3):137--145, 1995.
\newblock \href {http://arxiv.org/abs/q-alg/9502019}
  {\path{arXiv:q-alg/9502019}}, \href
  {http://dx.doi.org/10.1016/0370-2693(95)00386-Y}
  {\path{doi:10.1016/0370-2693(95)00386-Y}}.

\bibitem{Zakrzewski1997}
S.~Zakrzewski.
\newblock {Poisson Structures on the Poincar{\'{e}} Group}.
\newblock {\em Commun. Math. Phys.}, 185(2):285--311, 1997.
\newblock \href {http://arxiv.org/abs/q-alg/9602001}
  {\path{arXiv:q-alg/9602001}}, \href {http://dx.doi.org/10.1007/s002200050091}
  {\path{doi:10.1007/s002200050091}}.

\bibitem{BRH2003minkowskian}
A.~Ballesteros, N.~R. Bruno, and F.~J. Herranz.
\newblock {A non-commutative Minkowskian spacetime from a quantum AdS algebra}.
\newblock {\em Phys. Lett. B}, 574(3-4):276--282, 2003.
\newblock \href {http://arxiv.org/abs/hep-th/0306089}
  {\path{arXiv:hep-th/0306089}}, \href
  {http://dx.doi.org/10.1016/j.physletb.2003.09.014}
  {\path{doi:10.1016/j.physletb.2003.09.014}}.

\bibitem{BLT2016unified}
A.~Borowiec, J.~Lukierski, and V.~N. Tolstoy.
\newblock {Quantum deformations of D=4 Euclidean, Lorentz, Kleinian and
  quaternionic $\mathfrak o$*(4) symmetries in unified $\mathfrak o$(4; $\mathbb C$) setting}.
\newblock {\em Phys. Lett. B}, 754:176--181, 2016.
\newblock \href {http://arxiv.org/abs/1511.03653} {\path{arXiv:1511.03653}},
  \href {http://dx.doi.org/10.1016/j.physletb.2016.01.016}
  {\path{doi:10.1016/j.physletb.2016.01.016}}.

\bibitem{BLT2016unifiedaddendum}
A.~Borowiec, J.~Lukierski, and V.~N. Tolstoy.
\newblock {Addendum to ``Quantum deformations of D=4 Euclidean, Lorentz,
  Kleinian and quaternionic $\mathfrak o$*(4) symmetries in unified $\mathfrak o$(4; $\mathbb C$) setting"
  [Phys. Lett. B 754 (2016) 176?181]}.
\newblock {\em Phys. Lett. B}, 770:426--430, 2017.
\newblock \href {http://arxiv.org/abs/1704.06852} {\path{arXiv:1704.06852}},
  \href {http://dx.doi.org/10.1016/j.physletb.2017.04.070}
  {\path{doi:10.1016/j.physletb.2017.04.070}}.

\bibitem{MS2018constraints}
F.~Mercati and M.~Sergola.
\newblock {Physical constraints on quantum deformations of spacetime
  symmetries}.
\newblock {\em Nucl. Phys. B}, 933:320--339, 2018.
\newblock \href {http://arxiv.org/abs/1802.09483} {\path{arXiv:1802.09483}},
  \href {http://dx.doi.org/10.1016/j.nuclphysb.2018.06.014}
  {\path{doi:10.1016/j.nuclphysb.2018.06.014}}.

\bibitem{BM2018extended}
A.~Ballesteros and F.~Mercati.
\newblock {Extended noncommutative Minkowski spacetimes and hybrid gauge
  symmetries}.
\newblock {\em Eur. Phys. J. C}, 78:615, 2018.
\newblock \href {http://arxiv.org/abs/1805.07099} {\path{arXiv:1805.07099}},
  \href {http://dx.doi.org/10.1140/epjc/s10052-018-6097-1}
  {\path{doi:10.1140/epjc/s10052-018-6097-1}}.

\bibitem{Majid1988}
S.~Majid.
\newblock {Hopf algebras for physics at the Planck scale}.
\newblock {\em Class. Quantum Gravity}, 5(12):1587--1606, 1988.
\newblock \href {http://dx.doi.org/10.1088/0264-9381/5/12/010}
  {\path{doi:10.1088/0264-9381/5/12/010}}.

\bibitem{ChariPressley1994}
V.~Chari and A.~Pressley.
\newblock {\em {A guide to Quantum Groups}}.
\newblock Cambridge University Press, Cambridge, 1994.

\bibitem{Majid1995Book}
S.~Majid.
\newblock {\em {Foundations of quantum group theory}}.
\newblock Cambridge University Press, Cambridge, 1995.
\newblock \href {http://dx.doi.org/10.1017/CBO9780511613104}
  {\path{doi:10.1017/CBO9780511613104}}.
  
\bibitem{MajidCMS} S.~Majid.
\newblock {Meaning of Noncommutative Geometry and the Planck-Scale Quantum Group}.
\newblock {\em Towards Quantum Gravity}, J. Kowalski-Glikman (Ed.), 227--276. Springer Berlin Heidelberg, 2000.
\newblock  \href {https://doi.org/10.1007/3-540-46634-7_10}
  {\path{doi:10.1007/3-540-46634-7_10}}.


\bibitem{Kowalski-Glikman2013living}
J.~Kowalski-Glikman.
\newblock {Living in curved momentum space}.
\newblock {\em Int. J. Mod. Phys. A}, 28(12):1330014, 2013.
\newblock \href {http://arxiv.org/abs/1303.0195} {\path{arXiv:1303.0195}},
  \href {http://dx.doi.org/10.1142/S0217751X13300147}
  {\path{doi:10.1142/S0217751X13300147}}.

\bibitem{GM2013relativekappa}
G.~Gubitosi and F.~Mercati.
\newblock {Relative locality in $\kappa$-Poincar{\'{e}}}.
\newblock {\em Class. Quantum Gravity}, 30(14):145002, 2013.
\newblock \href {http://arxiv.org/abs/1106.5710} {\path{arXiv:1106.5710}},
  \href {http://dx.doi.org/10.1088/0264-9381/30/14/145002}
  {\path{doi:10.1088/0264-9381/30/14/145002}}.

\bibitem{BGGH2017curvedplb}
A.~Ballesteros, G.~Gubitosi, I.~Gutierrez-Sagredo, and F.~J. Herranz.
\newblock {Curved momentum spaces from quantum groups with cosmological
  constant}.
\newblock {\em Phys. Lett. B}, 773:47--53, 2017.
\newblock \href {http://arxiv.org/abs/1707.09600} {\path{arXiv:1707.09600}},
  \href {http://dx.doi.org/10.1016/j.physletb.2017.08.008}
  {\path{doi:10.1016/j.physletb.2017.08.008}}.

\bibitem{BGGH2018cms31}
A.~Ballesteros, G.~Gubitosi, I.~Gutierrez-Sagredo, and F.~J. Herranz.
\newblock {Curved momentum spaces from quantum (Anti-)de Sitter groups in (3+1)
  dimensions}.
\newblock {\em Phys. Rev. D}, 97(10):106024, 2018.
\newblock \href {http://arxiv.org/abs/1711.05050} {\path{arXiv:1711.05050}},
  \href {http://dx.doi.org/10.1103/PhysRevD.97.106024}
  {\path{doi:10.1103/PhysRevD.97.106024}}.

\bibitem{LSW2015hopfalgebroids}
J.~Lukierski, Z.~{\v{S}}koda, and M.~Woronowicz.
\newblock {$\kappa$-deformed covariant quantum phase spaces as Hopf
  algebroids}.
\newblock {\em Phys. Lett. B}, 750:401--406, 2015.
\newblock \href {http://arxiv.org/abs/1507.02612} {\path{arXiv:1507.02612}},
  \href {http://dx.doi.org/10.1016/j.physletb.2015.09.042}
  {\path{doi:10.1016/j.physletb.2015.09.042}}.

\bibitem{LMMPW2018algebroid}
J.~Lukierski, D.~Meljanac, S.~Meljanac, D.~Pikuti{\'{c}}, and M.~Woronowicz.
\newblock {Lie-deformed quantum Minkowski spaces from twists: Hopf-algebraic
  versus Hopf-algebroid approach}.
\newblock {\em Phys. Lett. B}, 777:1--7, 2018.
\newblock \href {http://dx.doi.org/10.1016/j.physletb.2017.12.007}
  {\path{doi:10.1016/j.physletb.2017.12.007}}.

\bibitem{Amelino-Camelia2001testable}
G.~Amelino-Camelia.
\newblock {Testable scenario for Relativity with minimum-length}.
\newblock {\em Phys. Lett. B}, 510(1-4):255--263, 2001.
\newblock \href {http://arxiv.org/abs/hep-th/0012238}
  {\path{arXiv:hep-th/0012238}}, \href
  {http://dx.doi.org/10.1016/S0370-2693(01)00506-8}
  {\path{doi:10.1016/S0370-2693(01)00506-8}}.

\bibitem{Kowalski-Glikman2001}
J.~Kowalski-Glikman.
\newblock {Observer-independent quantum of mass}.
\newblock {\em Phys. Lett. Sect. A Gen. At. Solid State Phys.},
  286(6):391--394, 2001.
\newblock \href {http://arxiv.org/abs/hep-th/0102098}
  {\path{arXiv:hep-th/0102098}}, \href
  {http://dx.doi.org/10.1016/S0375-9601(01)00465-0}
  {\path{doi:10.1016/S0375-9601(01)00465-0}}.

\bibitem{Amelino-Camelia2002planckian}
G.~Amelino-Camelia.
\newblock {Relativity in spacetimes with short-distance structure governed by
  an observer-independent (Planckian) length scale}.
\newblock {\em Int. J. Mod. Phys. D}, 11(1):35--59, 2002.
\newblock \href {http://arxiv.org/abs/gr-qc/0012051}
  {\path{arXiv:gr-qc/0012051}}, \href
  {http://dx.doi.org/10.1142/S0218271802001330}
  {\path{doi:10.1142/S0218271802001330}}.

\bibitem{MS2002}
J.~Magueijo and L.~Smolin.
\newblock {Lorentz Invariance with an Invariant Energy Scale}.
\newblock {\em Phys. Rev. Lett.}, 88(19):190403, 2002.
\newblock \href {http://arxiv.org/abs/hep-th/0112090}
  {\path{arXiv:hep-th/0112090}}, \href
  {http://dx.doi.org/10.1103/PhysRevLett.88.190403}
  {\path{doi:10.1103/PhysRevLett.88.190403}}.

\bibitem{LN2003versus}
J.~Lukierski and A.~Nowicki.
\newblock {Doubly special relativity versus $\kappa$-deformation of
  relativistic kinematics}.
\newblock {\em Int. J. Mod. Phys. A}, 18(1):7--18, 2003.
\newblock \href {http://arxiv.org/abs/hep-th/0203065v3}
  {\path{arXiv:hep-th/0203065v3}}, \href
  {http://dx.doi.org/10.1142/S0217751X03013600}
  {\path{doi:10.1142/S0217751X03013600}}.

\bibitem{BRH2003newdoubly}
A.~Ballesteros, N.~R. Bruno, and F.~J. Herranz.
\newblock {A new `doubly special relativity' theory from a quantum Weyl -
  Poincar{\'{e}} algebra}.
\newblock {\em J. Phys. A: Math. Gen.}, 36:10493--10503, 2003.
\newblock \href {http://arxiv.org/abs/hep-th/0305033}
  {\path{arXiv:hep-th/0305033}}, \href
  {http://dx.doi.org/10.1088/0305-4470/36/42/006}
  {\path{doi:10.1088/0305-4470/36/42/006}}.

\bibitem{FKS2004gravity}
L.~Freidel, J.~Kowalski-Glikman, and L.~Smolin.
\newblock 2+1 gravity and doubly special relativity.
\newblock {\em Phys. Rev. D}, 69(4):044001, 2004.
\newblock \href {http://arxiv.org/abs/hep-th/0307085}
  {\path{arXiv:hep-th/0307085}}, \href
  {http://dx.doi.org/10.1103/PhysRevD.69.044001}
  {\path{doi:10.1103/PhysRevD.69.044001}}.

\bibitem{ASS2004}
G.~Amelino-Camelia, L.~Smolin, and A.~Starodubtsev.
\newblock {Quantum symmetry, the cosmological constant and Planck-scale
  phenomenology}.
\newblock {\em Class. Quantum Gravity}, 21(13):3095--3110, 2004.
\newblock \href {http://arxiv.org/abs/hep-th/0306134}
  {\path{arXiv:hep-th/0306134}}, \href
  {http://dx.doi.org/10.1088/0264-9381/21/13/002}
  {\path{doi:10.1088/0264-9381/21/13/002}}.

\bibitem{Amelino-Camelia2010symmetry}
G.~Amelino-Camelia.
\newblock {Doubly-special relativity: Facts, myths and some key open issues}.
\newblock {\em Symmetry}, 2(1):230--271, 2010.
\newblock \href {http://arxiv.org/abs/1003.3942} {\path{arXiv:1003.3942}},
  \href {http://dx.doi.org/10.3390/sym2010230} {\path{doi:10.3390/sym2010230}}.

\bibitem{AFKS2011deepening}
G.~Amelino-Camelia, L.~Freidel, J.~Kowalski-Glikman, and L.~Smolin.
\newblock {Relative locality: A deepening of the relativity principle}.
\newblock {\em Gen. Relativ. Gravit.}, 43(10):2547--2553, 2011.
\newblock \href {http://arxiv.org/abs/1106.0313} {\path{arXiv:1106.0313}},
  \href {http://dx.doi.org/10.1007/s10714-011-1212-8}
  {\path{doi:10.1007/s10714-011-1212-8}}.

\bibitem{AFKS2011principle}
G.~Amelino-Camelia, L.~Freidel, J.~Kowalski-Glikman, and L.~Smolin.
\newblock {Principle of relative locality}.
\newblock {\em Phys. Rev. D}, 84(8):084010, 2011.
\newblock \href {http://arxiv.org/abs/1101.0931} {\path{arXiv:1101.0931}},
  \href {http://dx.doi.org/10.1103/PhysRevD.84.084010}
  {\path{doi:10.1103/PhysRevD.84.084010}}.

\bibitem{AAKRG2012relativelocality}
G.~Amelino-Camelia, M.~Arzano, J.~Kowalski-Glikman, G.~Rosati, and G.~Trevisan.
\newblock {Relative-locality distant observers and the phenomenology of
  momentum-space geometry}.
\newblock {\em Class. Quantum Gravity}, 29(7):075007, 2012.
\newblock \href {http://arxiv.org/abs/1107.1724v1} {\path{arXiv:1107.1724v1}},
  \href {http://dx.doi.org/10.1088/0264-9381/29/7/075007}
  {\path{doi:10.1088/0264-9381/29/7/075007}}.

\bibitem{HS1997phasespaces}
F.~J. Herranz and M.~Santander.
\newblock {Homogeneous phase spaces: the Cayley-Klein framework}.
\newblock In J.~F. Cari{\~{n}}ena, E.~Martinez, and M.~F. Ra{\~{n}}ada,
  editors, {\em Geom. y Fis. Memorias la Real Acad. Ciencias}, volume XXXII,
  pages 59--84, Madrid, 1998.
\newblock \href {http://arxiv.org/abs/physics/9702030}
  {\path{arXiv:physics/9702030}}.

\bibitem{Low1989}
R.~J. Low.
\newblock {The geometry of the space of null geodesics}.
\newblock {\em J. Math. Phys.}, 30(4):809--811, 1989.
\newblock \href {http://dx.doi.org/10.1063/1.528401}
  {\path{doi:10.1063/1.528401}}.

\bibitem{BP1991geodesics}
J.~K. Beem and P.~E. Parker.
\newblock {The space of geodesics}.
\newblock {\em Geom. Dedicata}, 38(1):87--99, apr 1991.
\newblock \href {http://dx.doi.org/10.1007/BF00147737}
  {\path{doi:10.1007/BF00147737}}.

\bibitem{Besse1978bookgeodesics}
A.~L. Besse.
\newblock {\em {Manifolds all of whose geodesics are closed}}, volume~93.
\newblock Springer Berlin Heidelberg, 1978.
\newblock
  \url{http://books.google.com/books?id=X8jn26cZWLQC{\%}5Cnpapers3://publication/uuid/2FA293ED-9DA4-4B78-8D61-4B75F899FDEB}.

\bibitem{AGK2011}
D.~V. Alekseevsky, B.~Guilfoyle, and W.~Klingenberg.
\newblock {On the Geometry of Spaces of Oriented Geodesics}.
\newblock {\em Ann. Glob. Anal. Geom.}, 40:389--409, 2011.
\newblock \href {http://arxiv.org/abs/0911.2602v1} {\path{arXiv:0911.2602v1}},
  \href {http://dx.doi.org/10.1007/s10455-011-9261-5}
  {\path{doi:10.1007/s10455-011-9261-5}}.

\bibitem{BRH2017}
A.~Ballesteros, N.~R. Bruno, and F.~J. Herranz.
\newblock {Non-commutative relativistic spacetimes and worldlines from 2+ 1
  quantum (anti) de Sitter groups}.
\newblock {\em Adv. High Energy Phys.}, 2017:7876942 (19 pages), 2017.
\newblock \href {http://arxiv.org/abs/hep-th/0401244}
  {\path{arXiv:hep-th/0401244}}, \href {http://dx.doi.org/10.1155/2017/7876942}
  {\path{doi:10.1155/2017/7876942}}.

\bibitem{BMN2017homogeneous}
A.~Ballesteros, C.~Meusburger, and P.~Naranjo.
\newblock {AdS Poisson homogeneous spaces and Drinfel'd doubles}.
\newblock {\em J. Phys. A: Math. Theor.}, 50(39):395202, 2017.
\newblock \href {http://arxiv.org/abs/1701.04902v1}
  {\path{arXiv:1701.04902v1}}, \href
  {http://dx.doi.org/10.1088/1751-8121/aa858c}
  {\path{doi:10.1088/1751-8121/aa858c}}.

\bibitem{Maslanka1993}
P.~Maslanka.
\newblock {The n-dimensional $\kappa$-Poincar{\'{e}} algebra and group}.
\newblock {\em J. Phys. A: Math. Gen.}, 26(24):L1251--L1253, 1993.
\newblock \href {http://dx.doi.org/10.1088/0305-4470/26/24/001}
  {\path{doi:10.1088/0305-4470/26/24/001}}.

\bibitem{NW1971}
T.~D. Newton and E.~P. Wigner.
\newblock {Localized states for elementary systems}.
\newblock {\em Rev. Mod. Phys.}, 21(3):400--406, jul 1949.
\newblock \href {http://dx.doi.org/10.1103/RevModPhys.21.400}
  {\path{doi:10.1103/RevModPhys.21.400}}.

\bibitem{BHOS1994global}
A.~Ballesteros, F.~J. Herranz, M.~A. del Olmo, and M.~Santander.
\newblock {Quantum (2+1) kinematical algebras: a global approach}.
\newblock {\em J. Phys. A: Math. Gen.}, 27(4):1283--1297, 1994.
\newblock \href {http://dx.doi.org/10.1088/0305-4470/27/4/021}
  {\path{doi:10.1088/0305-4470/27/4/021}}.

\bibitem{BHM2014plb}
A.~Ballesteros, F.~J. Herranz, and C.~Meusburger.
\newblock {A (2 + 1) non-commutative Drinfel'd double spacetime with
  cosmological constant}.
\newblock {\em Phys. Lett. B}, 732:201--209, 2014.
\newblock \href {http://arxiv.org/abs/1402.2884} {\path{arXiv:1402.2884}},
  \href {http://dx.doi.org/10.1016/j.physletb.2014.03.036}
  {\path{doi:10.1016/j.physletb.2014.03.036}}.

\bibitem{BHMN2017kappa3+1}
A.~Ballesteros, F.~J. Herranz, F.~Musso, and P.~Naranjo.
\newblock {The $\kappa$-(A)dS quantum algebra in (3 + 1) dimensions}.
\newblock {\em Phys. Lett. B}, 766:205--211, 2017.
\newblock \href {http://arxiv.org/abs/1612.03169} {\path{arXiv:1612.03169}},
  \href {http://dx.doi.org/10.1016/j.physletb.2017.01.020}
  {\path{doi:10.1016/j.physletb.2017.01.020}}.

\end{thebibliography}
\end{document}